\documentclass[letterpaper, amsfonts, amssymb, amsmath, reprint, showkeys, nofootinbib]{revtex4-1}

\usepackage{graphicx}
\usepackage{xcolor}
\usepackage{lipsum}
\usepackage{natbib}
\usepackage{physics}
\usepackage{vwcol}
\usepackage{hyperref}
\usepackage{upgreek}

\renewcommand{\mu}{\upmu}

\hypersetup{
    colorlinks = true,
    allcolors = black
}

\graphicspath{{figs/}}

\newcommand{\beginsupplement}{%
    \setcounter{table}{0}
    \renewcommand{\thetable}{S\arabic{table}}%
    \setcounter{figure}{0}
    \renewcommand{\thefigure}{S\arabic{figure}}%
    \renewcommand{\thesubsection}{\arabic{subsection}}
 }
 
\begin{document}
\title[Christen et al]{An integrated photonic engine for programmable atomic control}

\author{
Ian~Christen$^{1,*}$, 
Madison~Sutula$^1$, 
Thomas~Propson$^1$,
Hamed~Sattari$^2$,
Gregory~Choong$^2$,
Christopher~Panuski$^1$,
Alexander~Melville$^3$,
Justin~Mallek$^3$,
Scott~Hamilton$^3$,
P.~Benjamin~Dixon$^3$,
Adrian~J.~Menssen$^1$,
Danielle~Braje$^3$,
Amir~H.~Ghadimi$^{2,*}$, and
Dirk~Englund$^{1,}$}
\email{ichr@mit.edu, amir.ghadimi@csem.ch, englund@mit.edu}
\address{
$^1$Research Laboratory of Electronics, Massachusetts Institute of Technology, Cambridge, MA 02139, USA \\
$^2$Centre Suisse d’Electronique et de Microtechnique (CSEM), 2000 Neuch\^{a}tel, Switzerland \\
$^3$Lincoln Laboratory, Massachusetts Institute of Technology, Lexington, MA 02421, USA
}

\date{\today}

\begin{abstract}
Solutions for scalable, high-performance optical control are important for the development of scaled atom-based quantum technologies.
Modulation of many individual optical beams is central to the application of arbitrary gate and control sequences on arrays of atoms or atom-like systems.
At telecom wavelengths, miniaturization of optical components via photonic integration has pushed the scale and performance of classical and quantum optics far beyond the limitations of bulk devices~\cite{sun_2013_phased, poulton_2020_cmos, harris_2017_pnp}.
However, these material platforms for high-speed telecom integrated photonics~\cite{rahim_2018_Si, kish_2018_InP} are not transparent at the short wavelengths required by leading atomic systems~\cite{wan_2020_largescale, benhelm_2008_Ca, ebadi_2021_256}.
Here, we propose and implement a scalable and reconfigurable photonic architecture for multi-channel quantum control using integrated, visible-light modulators based on thin-film lithium niobate~\cite{desiatov_2019_visible, celik_2022_visible}.
Our approach combines techniques in free-space optics, holography, and control theory together with a sixteen-channel integrated photonic device to stabilize temporal and cross-channel power deviations and enable precise and uniform control.
Applying this device to a homogeneous constellation of silicon-vacancy artificial atoms in diamond, we present techniques to spatially and spectrally address a dynamically-selectable set of these stochastically-positioned point emitters.
We anticipate that this scalable and reconfigurable optical architecture will lead to systems that could enable parallel individual programmability of large many-body atomic systems, which is a critical step towards universal quantum computation on such hardware.
\end{abstract}


\maketitle

\section{Introduction}\label{intro}

\begin{figure*}%
\centering
\includegraphics[width=\textwidth]{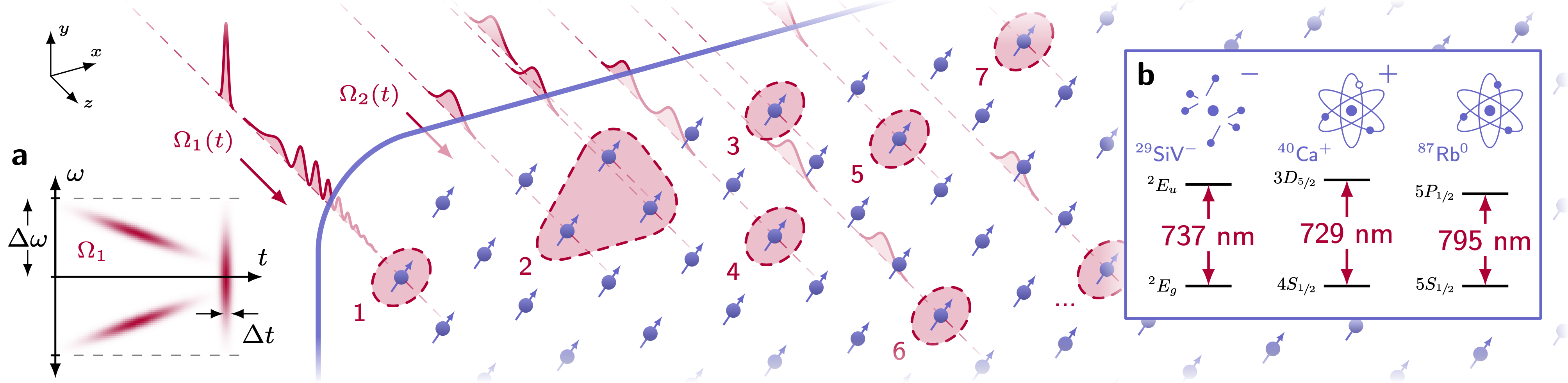}
\caption{
\textbf{Multi-channel optical addressing of atomic systems.}
Individually-controlled optical beams are incident upon single or grouped atomic systems, providing programmable control.
\textbf{a},
An example spectrogram of a waveform $\Omega_1$ consisting of a chirped pulse followed by a fast pulse incident on the channel labeled ``1''. Large switching bandwidth $\Delta\omega$ and, correspondingly, small switching time $\Delta t$ are desired for flexible spectro-temporal control.
\textbf{b},
Numerous atomic systems---from solid state memories, to ions, to neutral atoms---have visible wavelength transitions near 780~nm, a regime inaccessible by active photonics developed in silicon or indium phosphide for telecommunications.
We note a canonical example for each category, including especially the negatively-charged silicon-vacancy center in diamond which we use in Sec. \ref{diamond} to demonstrate the capabilities of our device.
}\label{fig1}
\end{figure*}
%
%

Controlling coherent light is essential to the use and understanding of atomic systems~\cite{wineland_1978_cooling, ashkin_1986_opticaltrap, anderson_1995_bec, monroe_1995_2qubit}.
A number of features are desired in optical control:
the application of frequency shifting~\cite{ebadi_2021_256} or frequency-domain shaping~\cite{debroux_2021_snvoptical, levine_2021_dispersive};
the execution of fast operations, for instance to compensate for finite atom lifetime~\cite{bernien_2017_51};
and the precise delivery of these optical phase and amplitude profiles.
Together, these features comprise the ideal of spectro-temporal control over an optical mode, where light is manipulated across frequency and time with precision.
The switching bandwidth of an optical modulator defines the extent and speed at which a spectro-temporal waveform can be realized.

Large-scale programmable quantum information processing on atomic systems requires the implementation of spectro-temporal control on individual spatially-distributed optical modes corresponding to atomic sites~\cite{cong_2022_Rbcontrol} (Fig. \ref{fig1}).
Previous work demonstrating multi-channel atomic addressing has largely involved extending bulk acousto-optic (AO) technologies---limited to $\mathcal{O}(\text{GHz})$ switching bandwidth---to multiple spatial channels, whether by mapping frequency domain signals to spatial sites via AO deflectors~\cite{lee_2016_individual, omran_2019_cat,  kranzl_2021_individual, graham_2022_individual} or by arraying many AO modulators~\cite{debnath_2016_individual, wright_2019_individual}.
Site count and modulation speed for AO deflectors is currently limited by acoustic velocity and bandwidth, especially when used in a rasterized mode, and the frequency gradient present in AO-deflected patterns is problematic for frequency-sensitive protocols without additional complexity~\cite{graham_2022_individual}.
Arrayed bulk modulators face the problem of scaling beyond the tens of channels demonstrated and towards the thousands or more necessary to realize fault-tolerant quantum computation~\cite{fowler_2012_surfacecodes}.
The complexity of assembling such devices scales with channel count, making growth by these orders of magnitude impractical.
A similar challenge led to the development of integrated classical computing technologies which transcended the limitations of bulk electronics via miniaturization of components and parallelization of fabrication~\cite{moore_1975_law}.

Likewise, integrated photonics enables scales and capabilities surpassing bulk optics~\cite{sun_2013_phased, poulton_2020_cmos, harris_2017_pnp, rahim_2018_Si, kish_2018_InP, rickman_2014_siphotonics}.
However, photonic integration at visible wavelengths has largely been passive, in part because silicon nitride (SiN), the prevailing material for visible photonics, does not have a strong intrinsic electro-optic effect~\cite{blumenthal_2020_review}.
While methods to add fast active modulation to SiN or similar passive materials have been investigated, none combine large and broadband switching bandwidth (DC to few GHz) with small switching voltages ($<$5~V)~\cite{phare_2015_graphene, yong_2022_multipass, liang_2021_vis, dong_2022_piezosin, dong_2022_piezosin2, tian_2020_sinstiff}.
Low switching voltages are important for direct compatibility with scalable high-speed complementary metal-oxide semiconductor (CMOS) electronics operating at $\mathcal{O}(1~\text{V})$~\cite{johansson_2014_cmos}.

\begin{figure*}%
\centering
\includegraphics{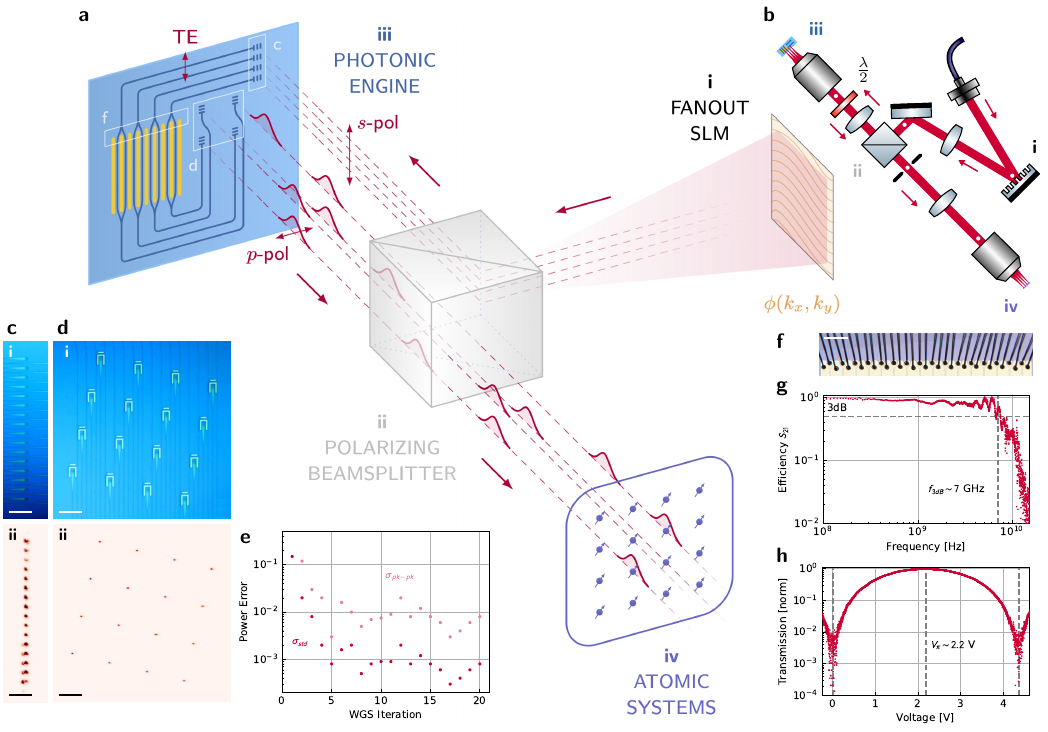}
\caption{
\textbf{Architecture, optical fanout, and modulator performance.}
\textbf{a},
An incident $s$-polarization hologram produced by a Fourier-domain SLM couples to many transverse electric (TE) waveguide modes via grating couplers (\textbf{i}-\textbf{iii}; \textbf{c}).
Integrated MZIs impart programmed waveforms on each channel. Output grating couplers direct light, now $p$-polarized, back to free-space towards the target sites (\textbf{iv}).
For better visibility, four channels---rather than the full sixteen---are shown in this diagram.
\textbf{b}, 
A top-down view of our setup illustrates the lenses and objectives omitted from the simplified diagram in \textbf{a}.
The half-waveplate in front of the objective aligns the beamsplitter and grating polarization axes.
A double-4$f$ lens configuration, with one lens participating in both 4$f$s, enables large fields of view. 
White dots represent imaging or Fourier planes.
\textbf{c}, 
The $1 \times 16$ input and \textbf{d}, the $4 \times 4$ output grating coupler arrays on our device in white light (\textbf{i}) and under coherent excitation (\textbf{ii}), showing the WGS-generated fanout hologram (\textbf{c ii}) and the resulting uniform output beams (\textbf{d ii}). Scalebars represent 50~$\mu$m.
\textbf{e}, Several iterations of WGS yield peak-to-peak optical output power errors $\sigma_\text{pk-pk}$ of approximately 1\% and standard deviations $\sigma_\text{std}$ of approximately 0.1\%.
\textbf{f},
Wirebonding to a PCB connects each of our sixteen channels to external control electronics. The scalebar represents 500~$\mu$m.
\textbf{g},
Switching bandwidth of a modulator channel.
\textbf{h},
A trace of device transmission versus voltage.
}\label{fig2}
\end{figure*}

\newpage
Thin-film lithium niobate (TFLN) has recently risen as an excellent platform for integrated nonlinear optics~\cite{wang_2018_lncmos, zhang_2019_comb, hu_2021_shift}.
TFLN combines large switching bandwidth and small switching voltage, with the state of the art pushing above 100 GHz~\cite{wang_2018_lncmos} and below one volt~\cite{ahmed_2020_subvolt}.
Moreover, lithium niobate possesses a wide bandgap with transparency down to 350~nm, allowing the visible-wavelength operation~\cite{desiatov_2019_visible, celik_2022_visible} which is critical for addressing many atomic systems.
Significant work over the past decade has made TFLN technologically ready to reliably fabricate large-scale circuits at wafer scales~\cite{luke_2020_waferscale, zhang_2021_review}, enabling us to focus on the system-level considerations which we resolve in this work:
\hyperref[device]{cross-channel uniformity}, 
\hyperref[stable]{precision stabilization}, and 
\hyperref[shaping]{reconfigurable projective mapping} between modulated channels and 
\hyperref[diamond]{atomic targets}.

\section{Architecture}\label{device}
Our photonic engine for optical quantum control consists of sixteen TFLN Mach-Zehnder interferometers (MZIs), routed to input ($1\times16$) and output ($4\times4$) grating banks at one side of the chip (Figs. \ref{fig2}c-d).
Each grating couples a waveguide mode at 780~nm into a free-space vertical Gaussian beam.
We measure channel insertion losses of approximately 20~dB at 780~nm, which we attribute to be dominated by grating inefficiency.
Each MZI amplitude modulator consists of two directional couplers and 3-mm-long phase shifters in push-pull ground-signal-ground configuration.
The MZIs share grounds and are wirebonded to a printed circuit board (PCB; Fig. \ref{fig2}f) for electrical control.
At 780~nm, we measure CMOS-compatible switching voltages of $V_\pi \sim 2.2$~V, extinction ratios exceeding 20~dB, and a PCB-limited 3~dB switching bandwidth of 7~GHz (Fig. \ref{fig2}g-h).

\begin{figure*}%
\centering
\includegraphics[width=\linewidth]{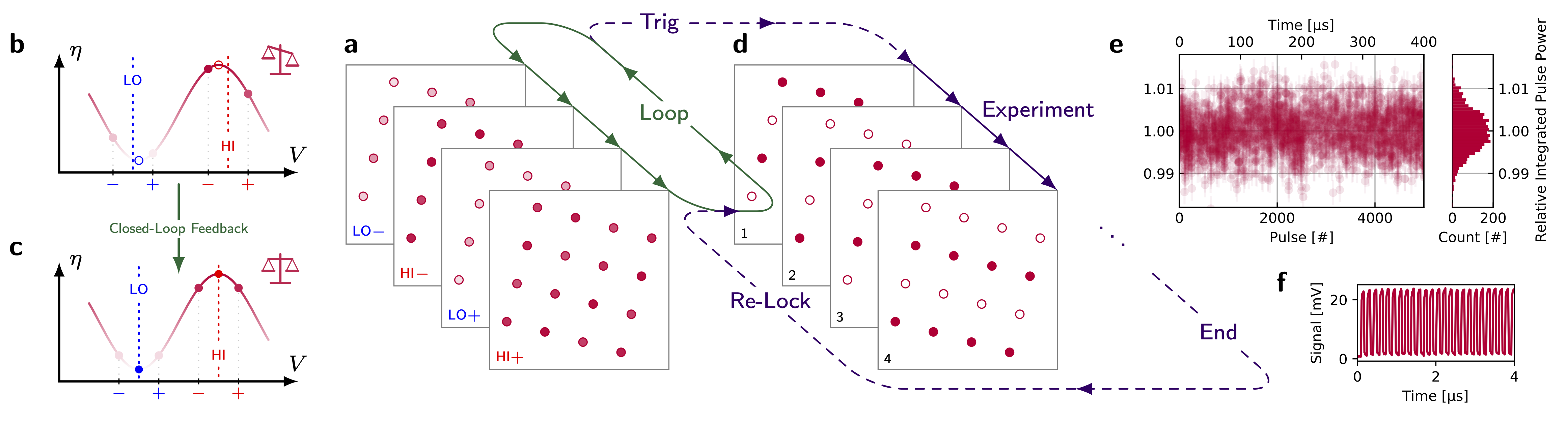}
\caption{
\textbf{Stabilizing zero-point drift.}
\textbf{a}, 
A camera-based feedback algorithm monitors and corrects the optoelectronic state of all channels by repeating measurements at four voltages.
\textbf{b},\textbf{c}, 
On each channel, we balance the measured modulator transmission $\eta$ of each voltage pair ($-,+$) and thus continuously refine alignment with the points of lowest (LO) and highest (HI) extinction.
\textbf{d},
Upon a hardware trigger, we lift the lock and apply desired waveforms to the modulators to control the target systems.
After completion, we reinstate the lock.
\textbf{e},
Integrated pulse powers from a sequence of square pulses on one channel; 1-$\sigma$ error derived from noise analysis is plotted.
\textbf{f},
A time trace of the first four microseconds of the same sequence.
}\label{fig3}
\end{figure*}
%
%

We use a low-bandwidth commercial liquid crystal on silicon (LCoS) spatial light modulator (SLM) to uniformly fanout optical power to our high-bandwidth integrated photonic channels.
This SLM produces a static hologram of beamspots coupled to the input grating couplers through the reflection port of a polarizing beamsplitter (Figs. \ref{fig2}a, \ref{fig2}c).
The orthogonal rotation of the output gratings relative to the input gratings (Fig. \ref{fig2}d) couples modulated light through the other port of the beamsplitter and towards the target atomic systems.
We monitor the output power of each channel with a camera and apply weighted Gerchberg–Saxton (WGS) feedback~\cite{dileonardo_2007_wgs} on the fanout hologram to refine cross-channel output uniformity, correcting for hologram alignment errors and variations in channel insertion loss.
A dozen iterations of WGS feedback yields uniformities better than 1\% (Figs. \ref{fig2}d ii, \ref{fig2}e).
While power fanout to each modulator could be accomplished monolithically with on-chip photonics or fiber-based splitters, the $\mathcal{O}(10^6)$ stable degrees of freedom in commercial SLMs permit a greater level of fanout precision and reliability.

\begin{figure*}%
\centering\includegraphics[width=1.025\textwidth]{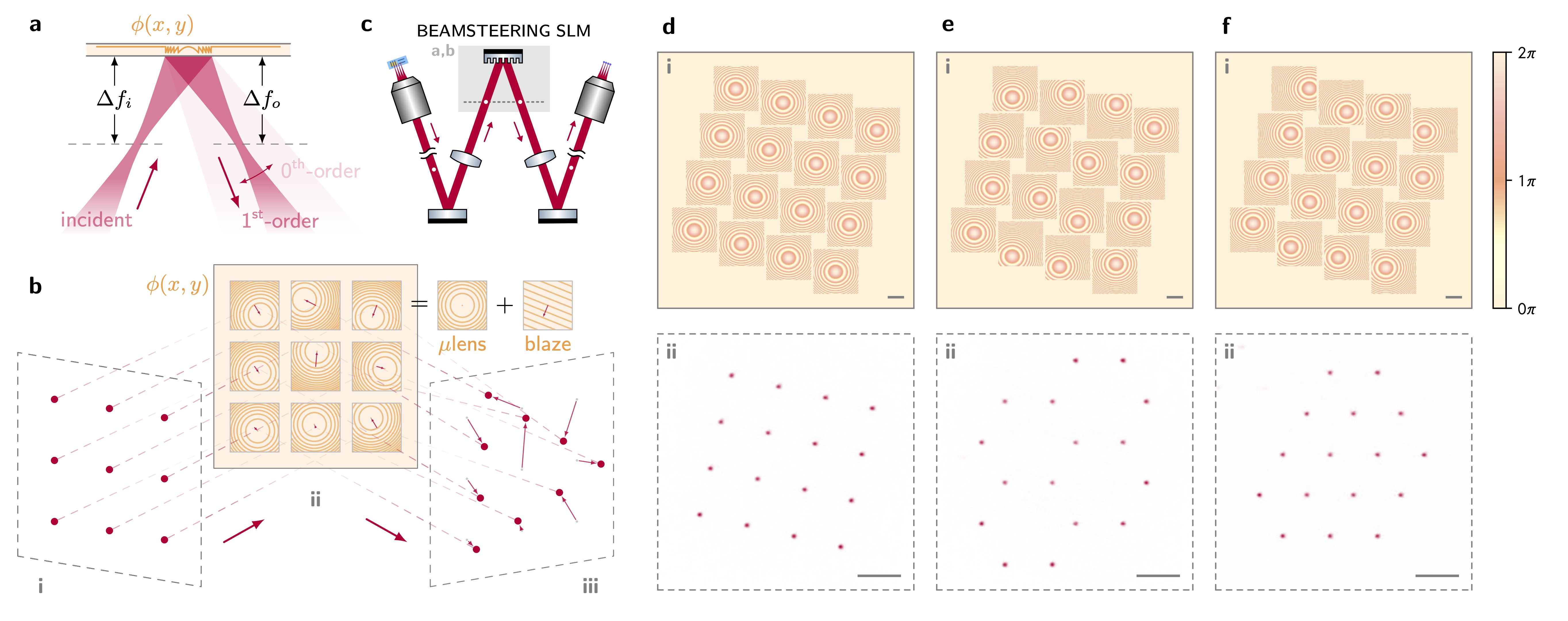}
\caption{
\textbf{Beamsteering output optical channels to target topologies.}
\textbf{a},
A microlens pattern $\phi(x,y)$ on a SLM which is defocused by $\Delta f_i$ is used to refocus incident diverging light to an output imaging plane defocused by $\Delta f_o$ (dashed lines).
The small zeroth-order reflection is defocused in this imaging plane, though this can be eliminated via Fourier domain filtering.
\textbf{b}, 
Adding a steering blaze to each microlens (\textbf{ii}) permits reconfiguration of the pattern, in this example from a grid (\textbf{i}) to a triangle (\textbf{iii}).
\textbf{c},
Our beamsteering optics positioned between the modulators (from left) and the target (to right).
\textbf{d},\textbf{e},\textbf{f},
Experimental demonstration of reconfiguration to example patterns: 
(\textbf{d}) square (no steering), 
(\textbf{e}) hexagonal, and 
(\textbf{f}) triangular lattices.
A microlens phasemask written on the SLM (\textbf{i}, top) produces each imaging domain pattern (\textbf{ii}, bottom).
A pickoff camera on the output path is used to capture these images, which were measured with $\Delta f_o \approx \Delta f_i$ at a wavelength of 737~nm.
Scalebars represent 200~$\mu$m at the SLM or camera (the camera has $\sim3\times$ magnification relative to the SLM).
Scaling the output patterns to micron-scale pitches is achievable with a shorter focal length imaging lens, as demonstrated in Fig. \ref{fig5}d.
}\label{fig4}
\end{figure*}

\section{Stabilization}\label{stable}
To stabilize the electrical degree of freedom of each optical channel, we developed a method for closed-loop parallel feedback on the sixteen arbitrary waveform generators controlling our channels.
Lithium niobate is known to be susceptible to zero-point drift, where the voltage corresponding to the point of highest extinction varies over time, which is attributed to fluctuations in charges trapped in the phase shifters~\cite{jiang_2017_drift, gozzard_2020_bulksteering, wang_2021_bulksteering}.
For bulk modulators, zero-point drift is mitigated via photodiode feedback and analog stabilization circuitry in commercial drivers.
Rather than arraying sixteen such drivers, we developed a CPU-based scheme that is scalable even beyond hundreds of channels.
Monitoring the optical channels with the same camera used for WGS optimization, we measure the transmission of each channel at four voltages: two pairs, each equally spaced around estimates for the modulator's minimum (LO) or maximum (HI) setpoint (Fig. \ref{fig3}a-b).
Closed-loop locking our setpoints to voltages with balanced pairwise transmission yields alignment with the locally-quadratic points of minimum or maximum transmission (Fig. \ref{fig3}c), while avoiding the challenge of directly measuring attenuated signals (i.e. LO) with short integration times.
We operate this locking loop at about 200~Hz, much faster than the $\mathcal{O}(\text{s})$ timescale of drift.

Desired waveforms or pulse sequences for atomic control are applied after a hardware trigger lifts the zero-point lock (Fig. \ref{fig3}d).
To characterize the performance of individual channels, we map output light to a fast photodiode and capture time traces.
Representative data from periodic square pulses is shown in Fig. \ref{fig3}e-f, demonstrating integrated pulse power deviations below 1\%.
For optical gates applied to quantum systems, this metric of integrated pulse power indirectly maps to phase accumulated or population driven during a gate pulse.
Repeatably achieving this performance from run-to-run is enabled by the quality of the closed-loop lock.
We observe zero-point drift on the time scale of several seconds when monitoring the drift of the modulators in a free-running open-loop configuration.
We find that this drift is negligible on the $\mathcal{O}(\text{ms})$ timescale of pulsesequences relevant to atomic experiments.
However, each time the closed-loop lock is lifted, we find that trajectory of zero-point drift is repeatable depending upon initial conditions, especially the average applied voltage in preceding seconds.

\section{Beamsteering}\label{shaping}
Next, we spatially reconfigure the output beam array of our device, which generally does not directly map to atomic targets.
In particular, the small array fill factor---the ratio of spot diameter to spot pitch---is not matched to 
densely packed atomic sites~\cite{graham_2022_individual}.
Furthermore, topological mismatch and beamline aberration limit the usefulness of directly projecting our square array of beams onto patterns of atomic systems.
For instance, ion crystals have increasingly non-uniform pitches towards their ends~\cite{debnath_2016_individual, wright_2019_individual}, and artificial atoms in solids are generally randomly distributed~\cite{sukachev_2017_siv, bersin_2019_multiplexing}.
Moreover, novel qubit connectivities for neutral atoms make use of free-form arrangements of optical traps~\cite{barredo_2018_3d, ebadi_2021_256, singh_2022_dual}.
Our optical control hardware achieves a level of reconfigurability that is able to address each of these cases, using a second LCoS SLM positioned in a defocused imaging plane.

Microlenses written to this SLM refocus the beams to an output plane (Figs. \ref{fig4}a-c).
Changing the defocusing distance $\Delta f_o$ of the output plane controls lattice fill factor, to the point of unity fill factor when the beams are each collimated by the microlenses ($\Delta f_o \rightarrow \infty$; see Methods Sec. \ref{fillfactor} and Fig. \ref{sup7}).
Linear phase gradients, equivalent to blazed gratings, added to each microlens steer the beams across the output plane, thus enabling reconfiguration of spatial spotpatterns to match desired topologies~\cite{ebadi_2021_256} (Figs. \ref{fig4}d-f).
The patterns are refined through WGS feedback on the input (fanout) SLM to recover uniformity lost to differing microlens efficiencies and other beampath distortions. 
We emphasize that each beamspot in each pattern is an individually-controllable optical degree of freedom which can be modulated with the high speed and precision demonstrated in sections \ref{device} and \ref{stable}.

Such a beamsteering scheme amounts to the free-form spatial reconfiguration of many fast temporal optical degrees of freedom, with applications from optogenetics~\cite{moreaux_2020_optogenetics} to optical ranging~\cite{zhang_2022_lidar} to augmented reality~\cite{xiong_2021_ar}.
In future upgrades, the demonstrated steering and fill factor conversion can be extended toward shaping unique holograms on each channel for enhanced beam uniformity or multi-site targeting~\cite{ebadi_2021_256, cong_2022_Rbcontrol}, as well as three-dimensional pattern generation~\cite{barredo_2018_3d}.
The reconfigurability of our architecture also permits pattern healing in the case of non-unity modulator yield, where non-functional channels can be covered by working channels to recover defect-free topologies, albeit with fewer total channels.

\section{Application}\label{diamond}

\begin{figure*}%
\centering
\includegraphics[width=\linewidth]{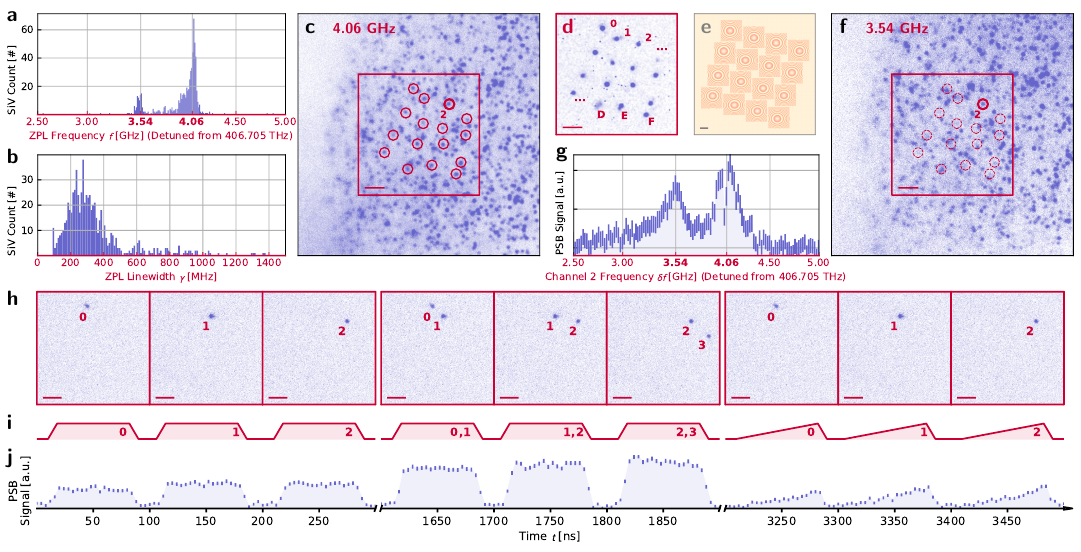}
\caption{
\textbf{Silicon-vacancy characterization and individual addressing.}
Population statistics for 953 emitters within our field of view: 
\textbf{a},
the frequency ($f$) distribution of the emitters, revealing two sharp spectral populations, and
\textbf{b},
the measured spectral linewidth distribution.
See extended data video 1 for the full three-dimensional data ($xyf$) used to produce these plots.
\textbf{c},
Spatial ($xy$) image of emitters resonant with $f_0=406.70906$~THz excitation. We select a subset of sixteen emitters (circled) which we
\textbf{d},
target with our modulator channels (labeled in hexadecimal; 0-F) using 
\textbf{e},
our beamsteering scheme, through the microlens phasemask pictured.
\textbf{f},
Spatial image of emitters resonant with a second color $f_1=406.70750$~THz. Channel 2 also has spatial overlap with an emitter at this second spectral plane.
\textbf{g},
A PLE fluorescence scan acquired by shifting the frequency of channel 2, showing peaks corresponding to the two targeted emitters.
\textbf{h},
Sections of a pulse sequence using our multi-channel device targeting different sets of emitters:
\textbf{i},
analog optical waveforms imparted upon the emitters yield 
\textbf{j},
globally-collected fluorescence.
Errorbars represent 1-$\sigma$.
Scalebars for \textbf{c}, \textbf{d}, \textbf{f}, and \textbf{h} represent 5~$\mu$m.
The scalebar for \textbf{e} represents 200~$\mu$m.
}\label{fig5}
\end{figure*}
%
%

With channel stability and reconfigurability established, we demonstrate the capabilities of our photonic engine using a layer of silicon-vacancy (SiV) color centers in mono-crystalline diamond cooled to 4~K~\cite{sukachev_2017_siv}.
To characterize this distribution of artificial atoms in space ($x$, $y$) and spectrum ($f$), we implement parallel-readout photoluminescence excitation (PLE) spectroscopy.
Exciting in widefield and collecting on a camera with single-emitter-level sensitivity, we scan the frequency of an incident laser and observe PLE signal on the frequency-detuned phonon sideband (PSB) when individual emitters are resonantly excited.
This process reveals an exceptionally narrow SiV spectral distribution for a solid-state system, with most emitters lying within two peaks which we attribute to orientation classes of emitters under global strain (Fig. \ref{fig5}a)~\cite{sutula_2022_widefield}.
We additionally measure most emitters to have linewidths close to the lifetime limit (Fig. \ref{fig5}b).
Setting the laser to the center of the denser distribution at 
$f_0 = 406.70906$~THz (737.11772 nm), we couple power through our modulators and steer the output beams to match the positions of sixteen isolated single emitters selected from the field (see Methods and Figs. \ref{fig5}c-e). 

One of these aligned beams, channel 2, has spatial overlap with a spectrally-detuned second emitter at $f_1 = f_0 + 520$~MHz (Fig. \ref{fig5}f).
Using channel 2, we demonstrate spectral addressability on two emitters by performing single channel PLE, made possible by the GHz-level switching bandwidth of our integrated modulators which exceeds the narrow spread of emitter frequencies in our sample.
We set the laser to $f_0-4.06$~GHz and frequency shift channel 2 by scanning a tone from 2.5 GHz to 5 GHz, observing the expected peaks at 3.54 and 4.06~GHz corresponding to the two emitters in channel 2 at $f_0$ and $f_1$ (Fig. \ref{fig5}g).
In this manner, a sample with wider or otherwise engineered inhomogeneous spectral distribution could be used to extend the number of emitters individually-addressable within a single field of view by a factor corresponding to the number of resolvable spectral sites~\cite{bersin_2019_multiplexing}.

In addition to spectral addressing, we implement a 5~$\mu$s-long pulse sequence for site-wise spatial addressing.
To measure these high speeds with efficiencies inaccessible to our camera, we couple the output emission to a free-space avalanche photodiode (APD), to map the sixteen emission spots to the active area of this detector.
The pulse sequence consists of three sections, each with sixteen pulses on a 100~ns period, demonstrating important concepts in multi-channel quantum optical control:
(1) individual emitter addressing,
(2) simultaneous addressing of multiple emitters---for example, towards pairwise entanglement generation---and
(3) analog tuning of the state of the modulators.
Figs. \ref{fig5}h-j contains results from channels 0 through 2 and pairs 0-1 through 2-3.
More information is displayed in Methods.

\section{Discussion}\label{discussion}
We have proposed and realized an architecture for scalable visible-wavelength optical programming, capable of high-speed control of many optical modes, each mode reconfigurable spatially and programmable across frequency and time.
Although this work has focused on the control of optical excitation, we emphasize the importance of collection optics which could also benefit from photonic integration. Collection and manipulation of single photons emitted from atomic systems could facilitate on-demand entanglement generation between targeted pairs of emitters or the construction of many-body nonclassical photonic states.
These collection schemes might take the form of spatial and spectral modulation of light collected into integrated waveguides~\cite{wan_2020_largescale} or free-space beamshaping, i.e. reciprocal to the integrated modulation and free-form beamsteering demonstrated on excitation light in this work.

Going forward, we see a number of paths to improve performance beyond these first-iteration designs.
For instance, grating efficiency will improve as design and fabrication matures, as realized in wafer-scale silicon photonics~\cite{vanvaerenbergh_2021_wafergrating}. 
Device extinction ratio, important for high fidelity quantum control, is likely limited by differences in splitting ratio between the two directional couplers which make up each MZI.
Avenues for improving extinction include optimizing coupler parameters and topology for increased tolerance to fabrication imperfections, or tuning couplers to recover performance.
At the cost of additional complexity, multiple MZIs could be cascaded per channel to reach exponentially-higher levels of extinction.
The compactness of current uncladded devices is constrained by a single in-plane layer of metal, which forces waveguides to be routed around large wirebond pads.
An additional out-of-plane layer of metal will allow us to approach closer to the limit of TFLN MZI compactness and realize substantially more channels.
While the footprint of these millimeters-long MZIs is much larger than that of resonant devices~\cite{liang_2021_vis}, channel count is ultimately limited by the length of a chip's perimeter in wirebonded control architectures, as the number of bondable electrical pads scales with perimeter.

Immediate scalability is an especially salient feature of these methods and this platform, enabled by integrated fabrication and switching voltages which can be directly synthesized by high-speed CMOS electronics.
To this end, the methods described in this work were investigated using dimensions compatible with much larger channel counts. 
For instance, only a fraction of available camera or beamsteering SLM area was used for the express purpose of field-testing focal lengths, image sizes, and superpixel dimensions suitable for operation with hundreds of channels.
In this manner, we anticipate that these advances in scalable optics will enable general quantum computation on the hundreds of coherent memories demonstrated in arrays of atomic systems~\cite{wan_2020_largescale, ebadi_2021_256}.

\appendix
\beginsupplement

\begin{figure*}%
\centering
\includegraphics[width=\textwidth]{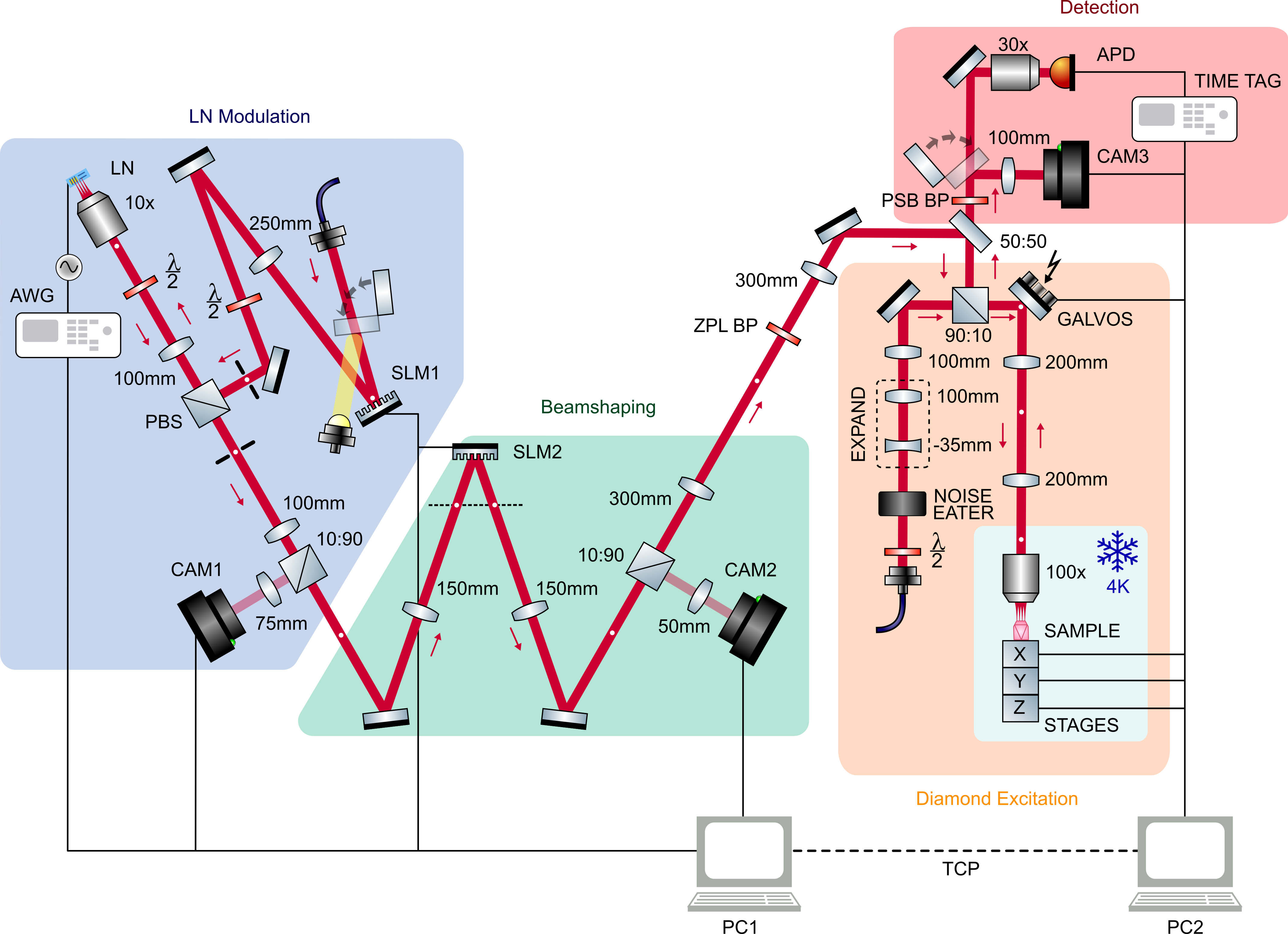}
\caption{
\textbf{Setup diagram,} summarizing components detailed in the main text along with Methods sections \ref{mod} and \ref{cyro}.
}\label{sup1}
\end{figure*}

\newpage

\section*{Methods}

\begin{figure*}%
\centering
\includegraphics{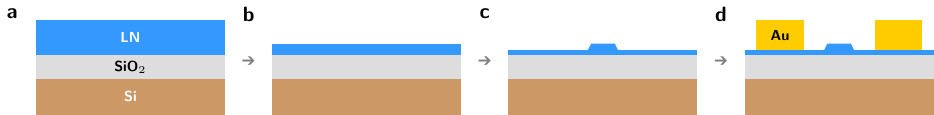}
\caption{
\textbf{Chip fabrication,} following steps detailed in Methods Sec. \ref{fab}.
}\label{sup2}
\end{figure*}

\subsection{Modulator Setup}\label{mod}
Fig. \ref{sup1} summarizes our optical setup. 
We collimate light from a polarization-maintaining fiber (Thorlabs PM780-HP) to a free-space beam with cm-scale diameter (Thorlabs ASL10142M-B; 79~mm asphere). 
White light (Thorlabs MWWHF2) is flipped into the path for diagnostic images.
The beam reflects off of the fanout SLM (SLM1; Thorlabs Exulus-HD3).
The plane of the SLM is scaled and mapped to the back aperture of the main objective (Zeiss Plan-NeoFluar, 10$\times$, .3~NA) via a 4$f$ (250~mm and 100~mm achromats), passing through the $s$~port of a polarizing beamsplitter (PBS; Thorlabs PBS252).
The hologram produced at the plane of the chip by SLM1 couples through each modulator channel and out the $p$~port of the PBS. 
A second 4$f$ (100~mm achromats), which shares a 100~mm lens with the first 4$f$, couples light to an intermediate Fourier plane.
From there, the light is partially reflected to a first pickoff camera (CAM1; Thorlabs Quantalux; 75~mm imaging achromat) to monitor the state of our channels.
The remaining light propagates further to the beamshaping setup.
The beamshaping SLM (SLM2; Santec SLM-200-01-0002-02) is positioned at a defocused imaging plane between two lenses (150~mm achromat).
A second pickoff camera (CAM2; Thorlabs Zelux; 50~mm imaging achromat) monitors the state of beamshaping.
From there, the light propagates to the cryostat setup described in Methods Sec. \ref{cyro}.

This setup, including the modulators and beamshaping optics, are controlled by a computer (PC1) with Python code.
Experiments that do not include color center addressing are controlled directly from this computer.
737~nm light is sourced from a Ti:Saph laser (MSquared SolsTiS) stabilized with a wavemeter (HighFinesse WS7; accuracy 60~GHz, precision 2~MHz).
For data taken at 780~nm, we use an external cavity diode laser (New Focus Velocity TLB-6712).
Each modulator is connected to an arbitrary waveform generator (AWG) channel sourced from four 4-channel cards (4 $\times$ Spectrum Instruments M2p.6566-x4, 16-bit, 125~Ms/s, 70~MHz).
For frequency shifting experiments, a faster AWG with only two channels (Tektronix AWG70002A, 25~Gs/s, 10~GHz) is used with a broadband amplifier (Centellax OA4MVM) to produce chirped sweeps.
Fast visible-wavelength photodiodes are used for high-speed measurements: for time traces (Melno Systems FPD510, 200~MHz) and bandwidth measurements (Newport 818-BB-45AF, 10~GHz). 
An oscilloscope (Agilent Infiniium DSO81004A, 10~GHz, 40~Gs/s) and a microwave vector network analyzer (Keysight N5224A PNA, 10~MHz to 43.5~GHz) are used to record signal from these photodiodes.

\subsection{Chip Fabrication}\label{fab}
Our chip was fabricated using a pre-commercial multi-project wafer (MPW) foundry service by CSEM.
The fabrication process flow begins with thinning down a $x$-cut lithium niobate on insulator (LNOI) thin film (NanoLN) from 600~nm thickness to 200~nm using blanket etching (Fig. \ref{sup2}a-b).
The 200~nm thickness is optimized for performance at wavelengths around 780~nm.
The resulting 200~nm LNOI thin-film is patterned using a HSQ mask via electron-beam lithography and etched a further 100~nm to yield low-loss optical TFLN waveguides with a sidewall slant approximately 30~degrees from normal (Fig. \ref{sup2}c).
For active electro-optic structures, 500~nm gold electrodes are patterned via liftoff (Fig. \ref{sup2}d).
We use 400~nm wide traces as standard single mode routing waveguides. 
These are tapered out slightly to 500~nm inside the phase shifters to reduce propagation loss. Phase shifters use a 3~$\mu$m gap between electrodes.
Waveguides are tapered down to 200~nm in directional couplers to increase the mode overlap between the two directional coupler arms.

\begin{figure*}%
\centering
\includegraphics{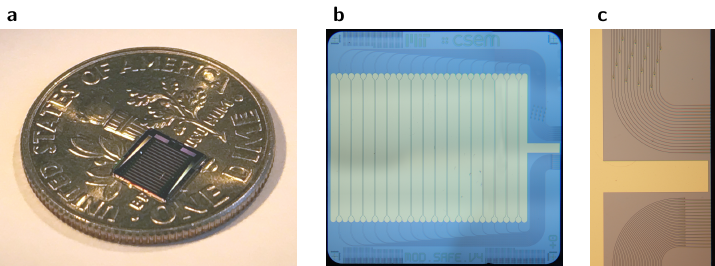}
\caption{
\textbf{Chip images.}
\textbf{a}, 
Our 5~mm~$\times$~5~mm chip imaged on top of a dime for scale and 
\textbf{b}, 
under an optical microscope.
\textbf{c}, 
Zoom upon the grating coupling region.
}\label{sup3}
\end{figure*}

\subsection{Chip Layout}\label{drop}
\begin{figure*}%
\centering
\includegraphics{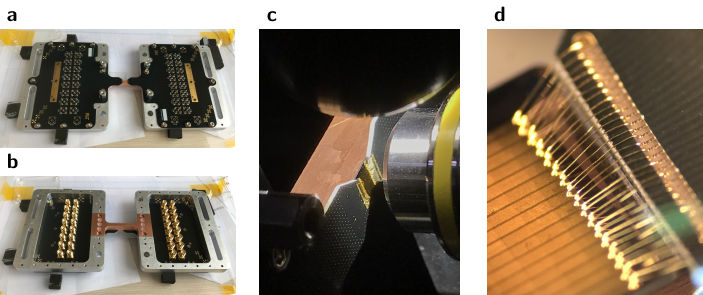}
\caption{
\textbf{Chip packaging.}
\textbf{a},\textbf{b}, The top and bottom of our chassis before chip placement. The chip sits at the center of copper beam in \textbf{a}. 
\textbf{c}, A wirebonded chip positioned in front of the objective.
\textbf{d}, Zoom upon an array of wirebonds.
}\label{sup4}
\end{figure*}

The photonic engine architecture described in the main text is fabricated on a 5~mm~$\times$~5~mm chip, consisting of a bank of sixteen MZIs with cross ports routed to banks of input and output grating couplers (Fig. \ref{sup3}).
The opposing two cross ports of each $2\times2$ MZI are routed to edge couplers via waveguides threaded inbetween the grating waveguides.
Our system could be operated with these edge couplers instead of the grating couplers.
In fact, our packaging was designed with this feature in mind, where the design of the chassis is compatible with a 90~degree rotation to expose the edge facet to objective imaging (Fig. \ref{sup4}a-c).
Edge-coupled operation is expected to perform with efficiencies closer to 2 dB channel insertion loss, given compatible optics and appropriate mode-matching.
However, in this work, we chose to focus on grating-coupled operation because of the greater generality of a two-dimensional grating-based architecture, especially with respect to arbitrary pattern generation and scalability.
Nevertheless, quantum systems with 1D topologies, such as linear ion crystals, are likely well suited to edge-coupled operation.

\subsection{Chip Packaging}\label{package}
The chip is glued with thermal epoxy (Arctic Silver) to a custom copper mount.
The copper mount is secured to custom aluminum parts, insulated with a layer of kapton tape and using teflon screws to enhance thermal isolation.
The copper mount is thermally stabilized with a thermoelectric cooler and temperature controller (Arroyo 5300 series).
The aluminum parts act as a mounting plate and support structure for custom FR4 PCBs which interpose sixteen SMA ports with the ground-signal-ground wirebond pads used to interface with our chip.
Together, these elements make up the chassis (Fig. \ref{sup4}a-b).
Chip and PCB pads are connected manually with a gold ball wirebonder with unity yield (Fig. \ref{sup4}c) allowing operation of every modulator.
Our bandwidth is likely limited by the roughly 50 mm long traces on our lossy FR4 PCB, along with contributions from wirebond length and compactness.

\begin{figure*}%
\centering
\includegraphics[width=.9\textwidth]{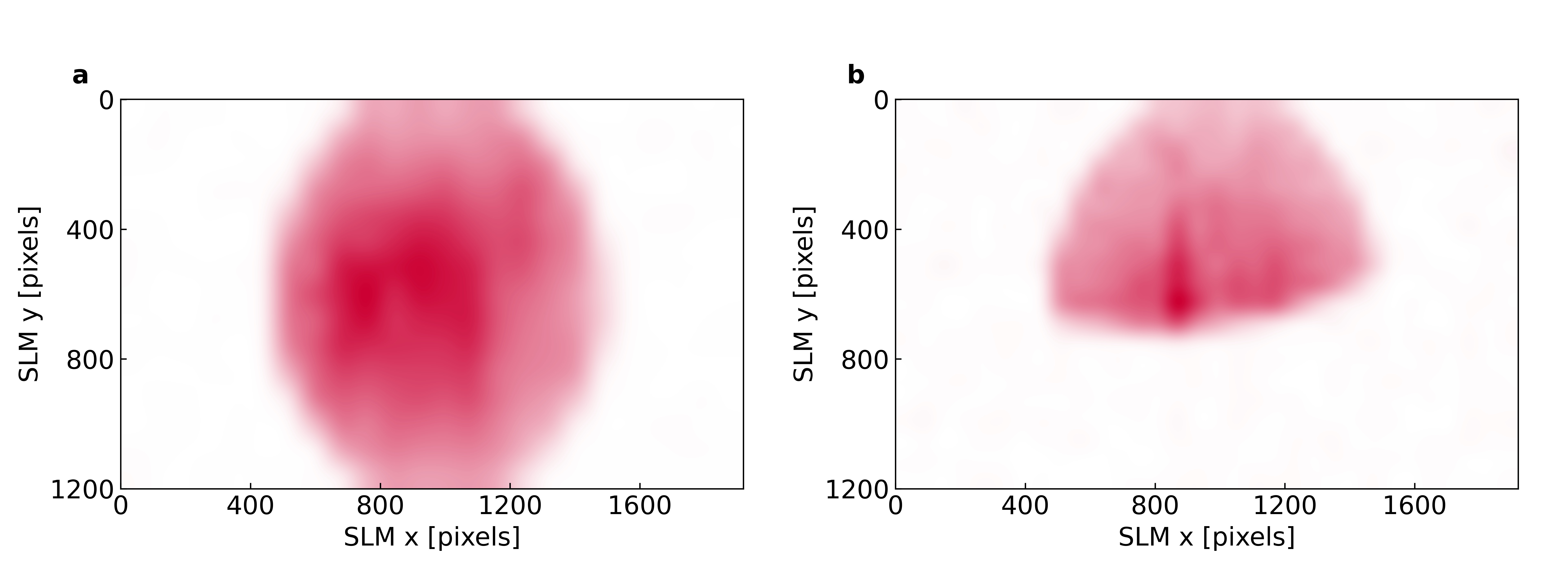}
\caption{
\textbf{SLM amplitude distributions.} 
The measured source amplitude distributions on the fanout SLM 
\textbf{a,} 
for 780~nm and 
\textbf{b,} 
for 737~nm.
}\label{sup5}
\end{figure*}

\subsection{Fanout Wavefront Calibration}\label{wavefront}
The generation of diffraction-limited holograms depends on accurate characterization and compensation of beampath aberrations.
We automatically measure the source laser's phase and amplitude distributions at the plane of the SLM via imaging domain interference of light diffracted from SLM superpixel clusters~\cite{cizmar_2010_calibration}.
Gerchberg–Saxton (GS) -type algorithms use the amplitude distribution when numerically generating phase profiles, enabling greater holographic accuracy.
We find that the amplitude distribution becomes significantly clipped as the reference position approaches the edge of the field of view.
To compensate for this, we purposefully overclip the SLM with an iris such that the amplitude distribution is equally clipped regardless of the reference position (clipping oval from the iris is visible on the amplitude distribution pictured in Fig. \ref{sup5}a).
This allows us to generate more accurate holograms across the region near the edge of the field of view where the input grating array is located.

\subsection{Grating Design, Coupling, and Efficiency}\label{wavefront}
The grating couplers are designed to couple light from a single mode waveguide to a vertical Gaussian beam with .12 numerical aperture (NA) at 780~nm.
Vertical coupling is important to remain within the NA of the normally-incident objective.
In the longitudinal direction---in the direction of propagation of the waveguide---we make use of gradient-based optimization~\cite{su_2018_spinsb}.
In the transverse direction, we use analytic assumptions of Gaussian phasefronts to shape the focus to that of the longitudinal direction.
We measure a slight ellipticity to the beam in waist and focus, corresponding to mismatch between the transverse and longitudinal directions, though this is compensated with the beamshaping microlenses.
Metal guards (visible as U shapes in Fig. \ref{fig2}di) are fabricated around the output gratings with the intent of reducing crosstalk scatter to neighboring outputs.

We engineered degrees of freedom in our setup to account for deviations from vertical grating coupling.
The mirror on the excitation path immediately before the polarizing beamsplitter is designed to be near an imaging plane (see white dot), such that changing the angle of this mirror directly tunes the angle of excitation at the sample's imaging plane to optimally couple the hologram into the gratings.
At 780~nm, we find that the optimal angle of excitation is close to normal and measure roughly 20~dB device insertion loss.
At 737~nm, we measure closer to 40 dB device insertion loss. 
We attribute this additionally loss to a non-vertical grating coupling angle at this non-design wavelength, finding evidence in the wavefront calibration for 737~nm, where the angle producing optimal coupling also causes significant amplitude distribution clipping (Fig. \ref{sup5}b).
We interpret this as clipping on the edge of the objective's NA.
More broadband gratings or gratings targeting 737~nm could be fabricated to avoid loss at this wavelength.

\begin{figure*}%
\centering
\includegraphics[width=\textwidth]{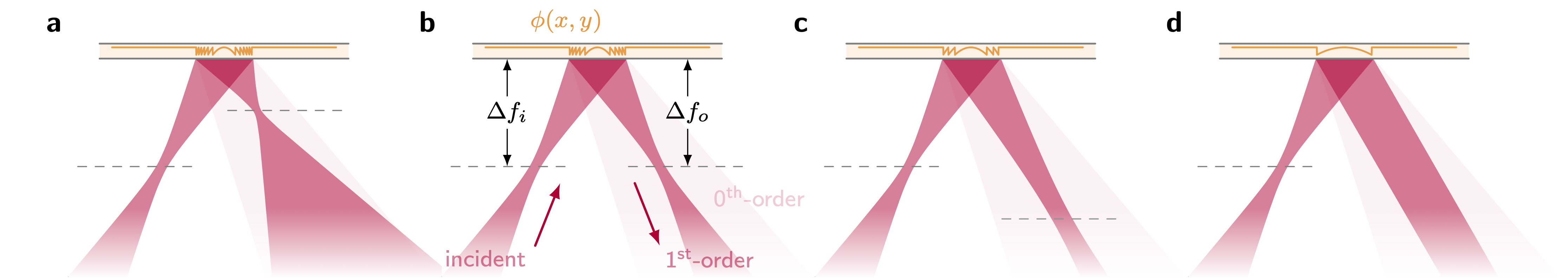}
\caption{
\textbf{Microlens fill factor conversion.}
The influence of a variety of $\Delta f_o$ upon the fill factor $\eta_o$ of the output beam:
\textbf{a},~$\Delta f_o = .5\Delta f_i$,
\textbf{b},~$\Delta f_o = \Delta f_i$,
\textbf{c},~$\Delta f_o = 1.5\Delta f_i$, and
\textbf{d},~collimated output.
}\label{sup7}
\end{figure*}
\begin{figure*}%
\centering
\includegraphics[width=.5\textwidth]{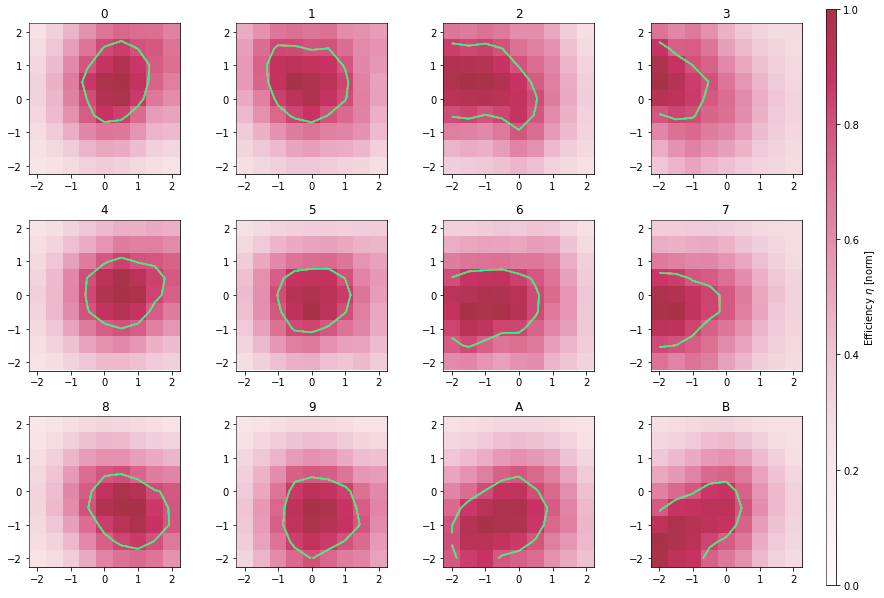}
\caption{
\textbf{Microlens steering range.}
Lens efficiency $\eta$ versus normalized steering range $R$ for twelve channels with $\Delta f_o \approx \Delta f_i$.
The $x$ and $y$ axes are normalized to the channel pitch $\Gamma$.
Contours denote 80\% efficiency.
}\label{sup8}
\end{figure*}

\subsection{Imaging Domain Calibration}\label{imaging_calibration}
We locate the positions of input and output gratings in imaging domain $x$-space manually using a reference image previously taken with white light.
The positions of the output gratings are used to center pixel integration regions about each output spot, which we use to determine the output power of each channel.
We also calibrate the coordinate transform between SLM $k$-space and the imaging domain of the chip.
This amounts to generating a grid pattern on the SLM and fitting the result to an affine transformation.

\subsection{Fanout Hologram Generation}\label{fanout}
For this section, a sawtooth electrical signal with peak-to-peak amplitude of $2V_\pi$ at a frequency of 100 cycles per camera frame is applied to every modulator.
This tone averages out fluctuations in device transmission due to the electrical degree of freedom, and isolates the problem of optimizing fanout coupling.
The inverse of the calibration coordinate transformation and $x$-space input grating positions found in Sec. \ref{imaging_calibration} are used to target positions in SLM $k$-space during GS fanout hologram optimization.
This first guess hologram is rarely perfectly aligned, which we attribute to chromatic aberration between the white light image and the target wavelength along with imperfections in our coordinate transform fitting.
To correct for this mismatch between guess and true $k$-vectors, we add a global steering blaze to the SLM and scan the position of the hologram across the gratings, measuring the coupling through the output gratings via camera readout.
We iteratively correct the $k$-vector guess for each channel by adding the blaze $k$-vector that produced maximal coupling in the previous iteration.
Four iterations yield alignment below measurement noise.

With the $k$-space map corrected, we use WGS to compensate for remaining pointing or device transmission errors. 
For the data presented in this work, we optimize the uniformity of integrated spot power, though other figures of merit such as spot amplitude uniformity are equally applicable for weighted optimization.
We believe that current uniformity is limited by small mechanical vibrations between the objective and chip causing pointing errors between the fanout hologram and input gratings.
This effect can be mitigated in future iterations by directly securing the objective to the chassis that supports the chip, rather than separately securing them upon the same optical breadboard.
Without any optical feedback or temperature stabilization, our system and holograms remain stable to within 10\% of baseline coupling over the course of 24 hours.

\begin{figure*}%
\centering
\includegraphics{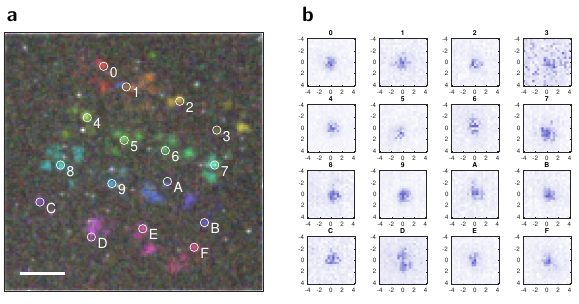}
\caption{
\textbf{Channel-emitter alignment.} 
\textbf{a,}
Composite blaze scans from each microlens (colored by hue) are used to select an isolated emitter at a target camera pixel for each channel (0-F). Scalebar represents 5~$\mu$m.
\textbf{b,}
After optimization in $x$, $y$, and focus, blaze scans collecting on the target pixel show centered flourescence.
The $x$ and $y$ axes use units of blaze angle at the beamsteering SLM in milliradians.
}\label{sup9}
\end{figure*}

\subsection{Closed-Loop Zero-Point Stabilization}
We preload the four-voltage locking sequence and desired pulse sequence to the AWGs.
The camera used to measure the state of each modulator at these voltages is triggered via a digital AWG line to maintain synchronization.
Updates to setpoints are applied to the ``amplitude'' and ``offset'' voltage parameters of the AWGs, which scale the unitless digital waveforms loaded to memory, avoiding time-consuming waveform memory rewriting.
We prioritize (via CPU-based analysis) updating channels that have the largest absolute error between the current setpoint and the measured target value, as updating these parameters is still slow (to the point that we can only update two parameters per camera frame).
Hardware without such update latency limitations can be used to eliminate this issue.

\subsection{Microlens Fill Factor Conversion}\label{fillfactor}
We define the fill factors $\eta_i$ and $\eta_o$ of the input and output beams as the ratios between the $1/e^2$ beamdiameters $2w_i$ and $2w_o$ and pitch $\Gamma$ of spots. For filled microlenses ($2w(z) = 2w(\Delta f) = \Gamma$) and an approximately diffraction-limited system, we estimate the defocusing distances $\Delta f_{i|o}$ between the SLM and the input and output planes to follow the relation:
\begin{align}
    \Gamma 
    = 2w_{i|o}\sqrt{1 + \left(\frac{\Delta f_{i|o} \lambda}{\pi w_{i|o}^2}\right)^2},
\end{align}
where $z_R^{i|o} = \pi w_{i|o}^2/\lambda$ is the Rayleigh range.
While the input defocusing distance $\Delta f_i$ is fixed by the choice of pitch $\Gamma$ and the fill factor $\eta_i$ of our system, we can engineer $\Delta f_o$ to target a desired $\eta_o$  (Fig. \ref{sup7}) according to:
\begin{align}
    \Delta f_o(\eta_o) &= \frac{\pi}{4}\frac{\Gamma^2 \eta_o}{\lambda}\sqrt{1 - \eta_o^2}.
\end{align}
Unity fill factor can be achieved when the output beams are collimated (Fig. \ref{sup7}d).

\subsection{Microlens Steering Range}
There are two main factors that limit steering range:
(i) the SLM's diffraction efficiency versus steered angle, and
(ii) the extent at which steering can be realized through high magnification objectives.

For commercial LCoS SLMs, diffraction efficiency degrades as target blaze gradients approach the point where they can no longer be resolved by finite pixel size.
The half-angular SLM steering bandwidth $\theta_\text{max}^\text{diff}$ is usually on the order of two degrees (the Bragg condition at 780~nm for a sawtooth blaze with a pitch of three 8~$\mu$m pixels).

Steering angles in the domain of the SLM are mapped to angles at the domain of the target, multiplied by a factor corresponding to the objective magnifcation $M$.
For the high ($M \sim 50\times$) objectives found in atomics experiments and given an appropriate imaging lens, a steering angle of two degrees at the SLM will not propagate through the objective as $M \times 2^{\circ} = 100^{\circ} > 90^{\circ}$ exceeds the NA of free-space.
A shallower angle may also clip upon the NA of the objective, bounded by $\theta_\text{max}^\text{obj} = \theta_\text{NA} / M$.
This limit could be completely negated with a third SLM used to reorient the angle of the beams to vertical incidence after they are spatially steered by the second SLM.

These bounds on steering angle pose a limit on $r$, the steering range, and $R$, the steering range normalized to pitch $\Gamma$:
\begin{align}
R = \frac{r}{\Gamma} < \frac{\theta_\text{max} \Delta f_o}{\Gamma} = \theta_\text{max} \frac{\pi}{4}\frac{\Gamma \eta_0}{\lambda}\sqrt{1-\eta_0^2}.
\end{align}
In our system, with $\Gamma \sim .65$~mm and $\lambda\sim780$~nm and using $\theta_\text{max} = \theta_\text{max}^\text{diff}$, this evaluates to roughly $R \sim 20\eta_o$.
For small $\eta_{i|o}$, the phase gradient of the parabolic focusing phase can also be large enough to locally exceed $\theta_\text{max}^\text{diff}$ on the edges of the lens, limiting $R$ further.
For the data presented in Fig. \ref{fig4}, we use $\eta_o \sim .05$.
We separately measure $R$ for this $\eta_o$, finding that steering within $R \sim 1$ maintains efficiencies over 80\%, in agreement with $R \sim 20\eta_0$ (Fig. \ref{sup8}).

Though we operate the beamshaping SLM at a roughly five degree reflection angle, this does not mitigate performance. That is, focal detunings resulting from this slant (e.g. at the edges of the microlens grid) do not exceed the Rayleigh length of the beams, for the input and output parameters which we consider.
Aberrations from these small focal detunings are corrected on a microlens-by-microlens basis via automated routines.

\subsection{Microlens Pattern Generation}
For the topologies displayed in Figure \ref{fig4}, we automatically steer beams to a target pattern defined in the beamshaping camera's (CAM2's) basis.
Channels are matched to target points by framing the task as a linear sum assignment problem.
Using knowledge of the focal lengths and defocusing distances, we analytically calculate and apply the blaze necessary to steer each channel to the target position.
Measuring the positional error between the steered position and the target, we iteratively apply this process until the channels are sufficiently aligned. Four iterations yield satisfactory alignment.

\begin{figure*}%
\centering
\includegraphics{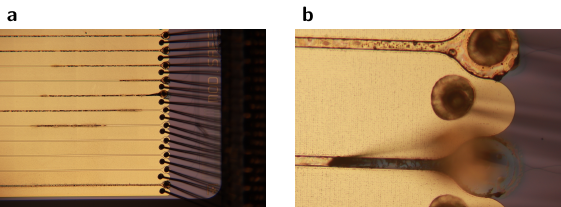}
\caption{
\textbf{Electrical breakdown damage.} \textbf{a,} Overview and \textbf{b,} zoom upon modulator charring and delamination which we attribute to an AWG over-voltage state.
}\label{sup11}
\end{figure*}
\begin{figure*}%
\centering
\includegraphics[width=\textwidth]{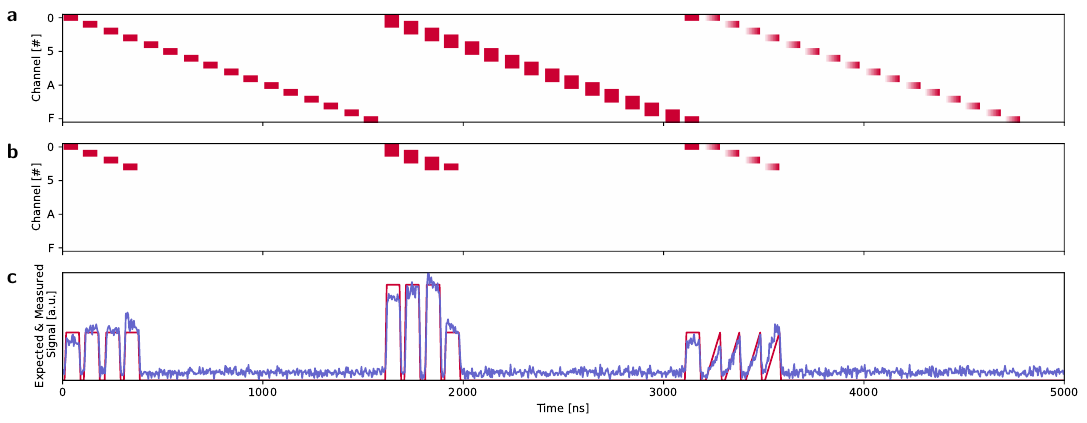}
\caption{
\textbf{Spatial Addressing Data.}
\textbf{a,}
Target pulse sequence on sixteen channels. Color represents channel amplitude for a given channel at a given time.
\textbf{b,}
Pulse sequence with our truncated channel count.
\textbf{c,}
Expected signal (red) as the sum of all globally-collected channels, compared with the measured fluorescence (blue).
}\label{sup10}
\end{figure*}

\subsection{Cryostat Setup}\label{cyro}
Fig. \ref{sup1} summarizes our optical setup.
Modulated and beamshaped light passes through a 4$f$ (300~mm achromats) from our modulator setup, passing through a cleanup filter (ZPL BP; Semrock BrightLine 740/13).
This merges with the excitation path of our cryostat setup with a removable 50:50 pellicle beamsplitter (Thorlabs BP145B2).
Galvos normally used for scanning confocal microscopy are used in this work for fine adjustments.
These galvos are mapped by a 4$f$ (200~mm achromats) to the back aperture of a cryo-optic objective (Zeiss 100$\times$, .95 NA) inside a 4K cryostat (Montana Instruments Cyrostation s50).
Our diamond sample is positioned under the objective with peizo slip stick stages (Attocube).
Emission light is collected back through the same path, to the other port of the pellicle.
Excitation light is filtered out, leaving the phonon sideband (PSB BP; Semrock Brightline 775/46).
From there, we collect on either a electron-multiplying charge-coupled-device (EMCCD) camera (CAM3; Photometrics Cascade 1K; 100~mm imaging achromat) or to an avalanche photodiode (APD; Excelitas Technologies SPCM-AQRH-14) with time-tagged signal (Swabian Instruments Time Tagger 20).
For widefield characterization data collected with the camera, we remove the pellicle and use a separate excitation port input on the 10 side of a 90:10 beamsplitter.
For this port, light is focused at the galvos (100~mm lens singlet), and thus---via the 4$f$---focused at the back aperture of the objective.
This focused beam in the Fourier domain of the objective maps to a widefield beam in the imaging domain.
A beam expander (-35~mm singlet, 100~mm singlet) increases the NA of the beam at the back aperture to fill the field of view in the imaging domain.
A compact noise eater (Thorlabs NEL03A) is used to stabilize power fluctuations occurring during laser wavelength sweeping.
Diamond experiments are controlled from a computer (PC2) which interfaces with the modulator computer (PC1) via an ethernet (TCP) link.
A mixture of MATLAB and Python code is used for these experiments.

\subsection{Silicon-Vacancy Sample}
Our diamond sample was produced by chemical vapor deposition (CVD) overgrowth on a low-strain mono-crystalline substrate (New Diamond Technologies).
Silicon is incorporated during the CVD process, yielding SiV color centers in the overgrown layer.
CVD overgrowth has previously been shown to generate narrow SiV inhomogeneous distributions similar to what we observe in this sample~\cite{rogers_2014_identical, sipahigil_2014_indistinguishable}.
Following overgrowth, the sample was cleaned by boiling in a 1:1:1 mixture of nitric, sulfuric, and perchloric acids for 1 hour~\cite{chu_2014_coherent}, cleaned in a 3:1 sulfuric acid:hydrogen peroxide piranha solution for 5 minutes, rinsed in a solvent bath, and annealed at 1200$^\circ$C in ultra-high vacuum.
To remove any graphite formed during annealing, the sample was again cleaned in tri-acid and piranha using the same processes.

\subsection{Channel-Emitter Alignment}
We adopt techniques from scanning microscopy, and use each microlens to scan each channel over a region of diamond (Fig. \ref{sup9}a). 
For each channel, we select a point in each scan corresponding to an isolated emitter.
Then, we iteratively optimize the blaze and focus of each microlens, maximizing camera signal corresponding to the chosen isolated emitter (Fig. \ref{sup9}b).

\subsection{Electrical Breakdown Damage}
After aligning our modulator channels to a set of sixteen SiV emitters, we proceeded to begin gathering final fluorescence data.
Despite working consistently for months beforehand, nine of our channels stopped coupling light during this acquisition (channels 4 through B and channels E through F).
We attribute this channel loss event to an untimely Windows Update producing an over-voltage state on the PCIe AWG cards which were active at the time, thus exceeding the electrical breakdown field of the air gap between the uncladded electrodes.
Resulting sparking across the gap accounts for the delaminated electrodes and charred waveguides which we observe on afflicted channels (Fig. \ref{sup11}).
As a result, the spatial addressing demonstration detailed in Fig. \ref{fig5} uses only the remaining modulator channels.

Breakdown can be nucleated by small sharp features which locally enhance electric field~\cite{slade_2002_breakdownsmallgap}.
The stochastic presence of such sharp features on each channel is a potential cause of the partial, rather than complete, destruction of our channels.
Future work using a silicon dioxide~\cite{austen_1940_breakdownglass} cladding will eliminate this problem as lithium niobate~\cite{luennemann_2003_breakdownln}, a stronger dielectric than air by more than an order of magnitude~\cite{davies_1965_breakdownair}, becomes the limiting material for electrical breakdown.
While this comparison is not strictly valid over the relevant electrodes gap sizes---due to differing breakdown mechanisms between solids and gasses along the with raised breakdown thresholds for small gaps~\cite{slade_2002_breakdownsmallgap}---the cited works illustrate the relative trends for these materials.
In cases where cladding is impractical, external overvoltage protection or alternative electronics could be used to prevent damage should a power outage or automated update occur.

\subsection{Spectral and Spatial Addressing}
The spectral plot Fig. \ref{fig5}g is acquired by chirping channel 2 at a frequency $\delta f$ using our 25~Gs/s AWG and collecting signal on our EMCCD camera.
Error is estimated from Poisson statistics.
Data is acquired with at 10~MHz steps, then binned to 20~MHz steps for visibility.
The choice of our sweep from 2.5 to 5~GHz stays within the bandwidth of our AWG and amplifier while avoiding spurious signal from higher order sidebands ($2\delta f$).

Our pulse sequence, as described in the main text, consists of a series of 80~ns pulses inside 100~ns bins.
Fig. \ref{fig5}h illusrates the spatial state of our modulators with camera images corresponding to a set of pulses, though these images are integrated for 5~seconds, instead of the 80~ns in the pulse sequence.
Fig. \ref{fig5}j was collected via repeating our pulse sequence over 20 minutes of integration.
Error is estimated from Poisson statistics.
The risetime slopes on Fig. \ref{fig5}i-j represent transitions from setpoint to setpoint of our 125~MS/s AWGs (8 ns per setpoint).
This risetime is nevertheless competitive with commercial AOMs.

\newpage

\section*{
Acknowledgments}
The authors thank 
Mikkel Heuck, 
Franco Wong, 
Artur Hermans, 
Alex Sludds, 
Saumil Bandyopadhyay, 
Matthew Trusheim, 
Eric Bersin,
Kevin Chen, 
Marco Colangelo,
Sara Mouradian, 
Xingyu Zhang,
Nathan Gemelke, 
Noel Wan, 
Frédéric Peyskens, 
Dongyuu Kim, 
Alex Lukin, 
Sepehr Ebadi, 
Dolev Bluvstein,
David Lewis,
and
Paul Gaudette
for enlightening discussions.

\section*{Declarations}

\subsection*{Funding}
I.C. acknowledges support from the National Defense Science and Engineering Graduate Fellowship Program and the National Science Foundation (NSF) award DMR-1747426. 
M.S. acknowledges support from the NASA Space Technology Graduate Research Fellowship Program and the NSF Center for Integrated Quantum Materials.
T.P. acknowledges support from the NSF Graduate Research Fellowship Program and the MIT Jacobs Presidential Fellowship.
C.P. acknowledges support from the Hertz Foundation Elizabeth and Stephen Fantone Family Fellowship.
A.J.M. acknowledges support from the Feodor Lynen Research Fellowship, the Humboldt Foundation, the MITRE Moonshot program, and the DARPA ONISQ program.
This work is supported by a collaboration between the US DOE and other Agencies. This material is based upon work supported by the U.S. Department of Energy, Office of Science, National Quantum Information Science Research Centers, Quantum Systems Accelerator (QSA).
This material is based upon work supported by the Air Force Office of Scientific Research under award number FA9550-20-1-0105, supervised by Dr. Gernot Pomrenke.
Experiments were supported in part by the NSF Center for Ultracold Atoms (CUA).
Lithium niobate fabrication at the CSEM was supported by the European Union’s Horizon 2020 research and innovation programme under grant agreement No. 101016138.
Distribution Statement A. Approved for public release. Distribution is unlimited. This material is based upon work supported by the National Reconnaissance Office (NRO) under Air Force Contract No. FA8702-15-D-0001. Any opinions, findings, conclusions or recommendations expressed in this material are those of the authors and do not necessarily reflect the views of the National Reconnaissance Office. \copyright 2022 Massachusetts Institute of Technology. Delivered to the U.S. Government with Unlimited Rights, as defined in DFARS Part 252.227-7013 or 7014 (Feb 2014). Notwithstanding any copyright notice, U.S. Government rights in this work are defined by DFARS 252.227-7013 or DFARS 252.227-7014 as detailed above. Use of this work other than as specifically authorized by the U.S. Government may violate any copyrights that exist in this work.

\subsection*{Competing Interests}
D.E. is a scientific advisor to and holds shares in QuEra Computing.
The CSEM offers lithium-niobate-on-insulator-related design, fabrication, testing, and integration services.


\subsection*{Availability of data and materials}
Available on request to I.C.

\subsection*{Code availability}
Available on request to I.C. SLM-related routines will be public on GitHub under the \verb+slmsuite+ package.

\subsection*{Authors' contributions}
I.C., A.J.M., and D.E. conceived the experiments.
I.C. designed systems, performed experiments, analyzed the data, and wrote the manuscript
with  assistance from T.P. and C.P. (software) along with M.S. and A.J.M. (experimental).
H.S., G.C., and A.G. fabricated TFLN devices.
M.S. characterized the diamond sample produced by A.M., J.M., S.H., P.B.D., and D.B.
D.E. and A.G. supervised the project.
All authors discussed the results and contributed to the manuscript.

\bibliography{bib/bib}

\end{document}